\documentclass[conference]{IEEEtran}
\IEEEoverridecommandlockouts
\pdfoutput=1
\usepackage{cite}
\usepackage{amsmath,amssymb,amsfonts}
\usepackage{algorithmic}
\usepackage{graphicx}
\usepackage{hyperref}
\usepackage{textcomp}
\usepackage{xcolor}
\def\BibTeX{{\rm B\kern-.05em{\sc i\kern-.025em b}\kern-.08em
    T\kern-.1667em\lower.7ex\hbox{E}\kern-.125emX}}
\begin{document}

\title{MOTIVE: Micropayments for trusted vehicular services\\
}

\author{\IEEEauthorblockN{Gowri Sankar Ramachandran, Xiang Ji, Pavas Navaney, Licheng Zheng, Martin Martinez, and\\ Bhaskar Krishnamachari}
\IEEEauthorblockA{\textit{USC Viterbi School of Engineering, University of Southern California, Los Angeles, USA} \\
\{gsramach, xiangji, navaney, lichengz, mart698, bkrishna\}@usc.edu}
}

\maketitle

\begin{abstract}
Increasingly, connected cars are becoming a decentralized data platform. With greater autonomy, they have growing needs for computation and perceiving the world around them through sensors. While today’s generation of vehicles carry all the necessary sensor data and computation on-board, we envision a future where vehicles can cooperate to increase their perception of the world beyond their immediate view, resulting in greater safety, coordination and more comfortable experience for their human occupants. In order for vehicles to obtain data, compute and other services from other vehicles or road-side infrastructure, it is important to be able to make micro-payments for those services and for the services to run seamlessly despite the challenges posed by mobility and ephemeral interactions with a dynamic set of neighboring devices. We present MOTIVE, a trusted and decentralized framework that allows vehicles to make peer-to-peer micropayments for data, compute and other services obtained from other vehicles or road-side infrastructure within radio range. The framework utilizes distributed ledger technologies including smart contracts to enable autonomous operation and trusted interactions between vehicles and nearby entities.
\end{abstract}


\section{Introduction}
The automotive industry is moving towards autonomous driving~\cite{wachenfeld2016use}, connectivity~\cite{gerla2014internet}, and electrification~\cite{mwasilu2014electric} to increase driver comfort, safety and eco-friendly driving. Modern day vehicles are increasingly being equipped with sensors, cameras, artificial intelligence, and machine learning algorithms to assist the drivers, or in some cases, the technologies are making decisions for the drivers. The ongoing developments in artificial intelligence technologies are expected to further propel the adoption of such technologies in the race for the realization of connected and autonomous vehicles. Thus, applications such as electric vehicle charging and real-time traffic prediction are expected to exchange data and computation using vehicle-to-vehicle (V2V) and vehicle-to-infrastructure (V2I) communication~\cite{siegel2018survey}.

Connected vehicles in the future can communicate with other vehicles on the road to help the driver make informed decisions when crossing intersections, changing lanes, or driving under adverse weather conditions. In such cases, each vehicle can exchange information such as the speed and their distance with other vehicles, among other things, to calculate the safe driving speed and distance. Similarly, the vehicles can communicate and coordinate with road-side infrastructure to navigate through intersections by controlling their speed based on the traffic conditions and the real-time local map. Such applications require aggregation and processing of sensor data from multiple vehicles and the road-side units. In summary, the V2V and V2I (referred to as V2X from now on) applications require communication and computation services from nearby vehicles and road-side units in a dynamic environment.

The V2X applications exchange data and compute services between devices and infrastructures owned by multiple users and organizations under transient conditions. In such scenarios, a trust mechanism is necessary to consume and provide services in the V2X application domain as the vehicles and the drivers may be misguided by the dishonest devices since the interactions between devices cross the trust boundaries. Besides, the introduction of micropayments encourage vehicles and infrastructure nodes to contribute resources including data and computation, allowing them to gain monetary benefit for their service contributions. It is essential to create a V2X platform with built-in mechanisms to guarantee trust and manage micropayments for exchanging data and compute services. Such a V2X platform has the potential to enable what we call ``financially autonomous vehicles".  Another problem in the V2X application scenario is that each vehicle stays in contact with other vehicles or the road-side units for a short amount of time. Thus, the service agreements between devices have to be made based on the contact duration for reliable transaction.

 In this paper, we introduce MOTIVE (an acronym coined from "Micropayments fOr Trusted vehIcular serVicEs"), a novel V2X platform with support for trust management, micropayments, and mechanisms to provide and consume data and compute services with other vehicles and road-side units following a decentralized architecture. MOTIVE incorporates a link prediction algorithm which allows the vehicles to calculate the contact duration based on the destination of the vehicles, speed, and the traffic conditions of the environment. MOTIVE is a blockchain and protocol agnostic framework for V2X involving blockchain and distributed ledger technologies. We list the open challenges and design trade-offs that are worthy of further research towards the realization of micropayments for V2X applications. A preliminary implementation shows the practicality of the proposed approach.

Section~\ref{sec:rw} discusses the related work. The architecture of MOTIVE is presented in Section~\ref{sec:arch}. Research challenges and design choices of MOTIVE are presented in Section~\ref{sec:rc}. Section~\ref{sec:pi} discusses our preliminary implementation of this framework. Finally, we conclude the paper with pointers for future work in Section~\ref{ref:con}.

\section{Related Work}
\label{sec:rw}
The application of blockchain to V2X are already discussed in the literature. ChargeItUp~\cite{Pedrosa:2018:CBT:3211933.3211949} uses micropayments for fuel recharging using layer-2 state channel mechanism. Leiding \emph{et al.}~\cite{Leiding:2016:SBV:2968219.2971409} present an architecture for a self-managed blockchain-based platform for V2X based on smart contracts. These works illustrate the benefits of using blockchain and distributed ledger technologies to V2X applications. The Chorus framework~\cite{leiding2018enabling} also explains the benefits of applying blockchain to V2X applications while contributing a technical architecture illustrating the components of the blockchain-based platform for V2X applications. Our work is complementary to~\cite{leiding2018enabling}~\cite{Leiding:2016:SBV:2968219.2971409}, except that the MOTIVE framework is more generic and it consists of components of link prediction and trust management. Zhaojun \emph{et al.} contribute BARS~\cite{8455893}, a trust management platform for V2X, which is one of the critical challenges in decentralized applications. BARS manages and estimate trust rating based on the historical behavior of the users and the indirect information collected from other vehicles in the vicinity. 

\textbf{Communication:} The 5G Automotive Association (5GAA) is standardizing 5G cellular technology for V2X and ITS applications~\cite{5gaa}, which is expected to advance connected vehicle technologies. A secure communication scheme for V2X applications is presented in~\cite{8442848}, which proposes a ring-based signature scheme in combination with multi-party smart contracts to ensure secure data sharing in a V2X environment. Merlin~\cite{bhargava2016optimizing} contributes a file sharing protocol for intermittently connected vehicles in V2X applications. Ahn~\emph{et al.}~\cite{6573913} contribute a file dissemination protocol for dual radio vehicular platform, which calculates the optimal content dissemination model for V2X applications. The literature on communication~\cite{5gaa,8442848} and data dissemination~\cite{bhargava2016optimizing,6573913} emphasizes the need to manage the data and service exchanges within the short contact duration. To these, our work contributes a novel link prediction algorithm as well as the use of trust-based micropayments.


\section{MOTIVE Architecture}
\label{sec:arch}
\begin{figure}[b]
\begin{center}
    \centerline{\includegraphics[height=3in,width=3in, keepaspectratio]{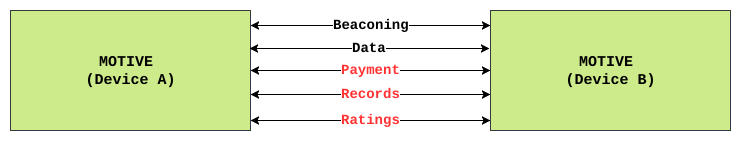}}
    \caption{Channels between MOTIVE Instances.}
    \label{fig:channels}
\end{center}
\end{figure}

\begin{figure}[t]
\begin{center}
    \centerline{\includegraphics[height=3in,width=3in, keepaspectratio]{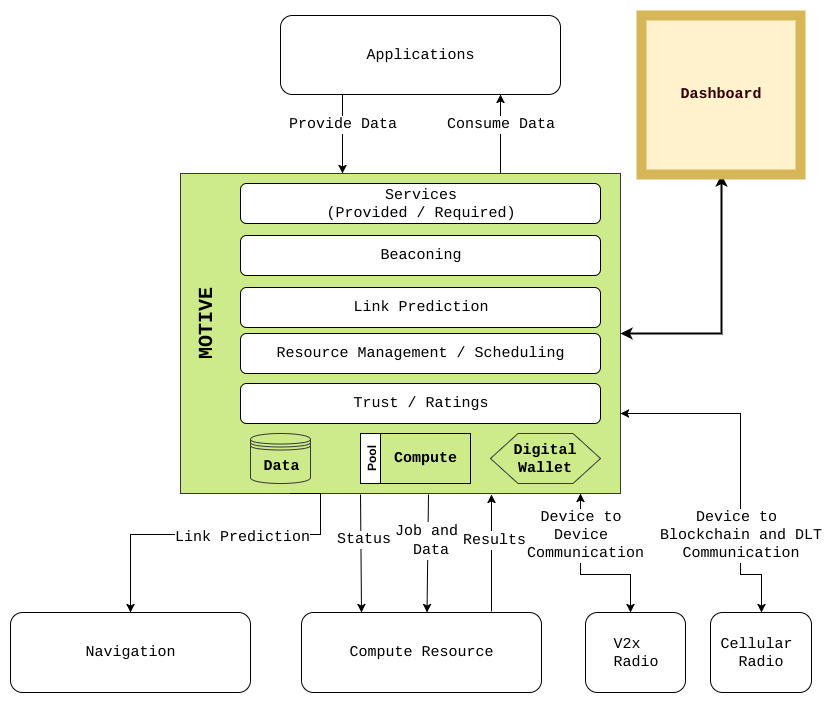}}
    \caption{Architecture of MOTIVE.}
    \label{fig:architecture}
\end{center}
\end{figure}
MOTIVE enables vehicles and infrastructures to exchange data and compute services in a peer-to-peer network, with built-in support for trust and micropayments. Figure~\ref{fig:channels} shows the different communication channels that exist between two MOTIVE instances. Each MOTIVE instance is:
\begin{itemize}
    \item Capable of \textbf{beaconing} the provided and required data and compute services to other MOTIVE instances in the neighborhood using a V2X radio.
    \item Able to exchange \textbf{data}, including sensor reading, computation tasks, inputs, and results, with other MOTIVE instances.
    \item Storing the \textbf{records} of all the transactions in a distributed ledger or a blockchain platform.
    \item Attached to a digital wallet for paying or receiving \textbf{payments} for the services consumed or provided respectively.
    \item Able to \textbf{rate} the other MOTIVE instance based on the transaction experience. 
\end{itemize}

\subsection{System Description}
Figure~\ref{fig:architecture} shows the architecture and the building blocks of MOTIVE. Each MOTIVE instance is capable of advertising the services it can provide and the services it requires to other MOTIVE instances using the V2X radio. Note that the MOTIVE instances can be running on both the vehicles and the road-side infrastructure. 

All the MOTIVE devices in the operational environment can receive beacons and process the information in the beacon to understand the provided and required services. The decision to offer the service to the other devices are made based on how long the two devices will stay in contact along with the rating and the account balance in the wallet. MOTIVE consists of a link prediction algorithm, which computes the contact duration based on the navigation data, traffic situation, distance between the vehicles, wireless communication range of the radio, and the speed limit. When the contact duration is well within the duration required to complete the transaction, the reputation of the device is above the acceptable threshold, and the account has sufficient balance to pay for the requested service, the service agreement is made between devices. The service provider then schedules the services by allocating the desired computation and storage resources and serve the peer in return for micropayment. The key building blocks of MOTIVE are described below:

\textbf{Beaconing:} Each MOTIVE instance advertises the services provided and required through a V2X radio. The device beacons as long as it has resources to provide and consume services. \textbf{Services:} Each MOTIVE instance can provide and consume data and computation services. \textbf{Data services} include information such as the number of vehicles in the vicinity or the electric charge available in the charging stations in a freeway. For example, a speed recommendation application can consume data from other cars in the neighborhood to estimate the desired speed for the current road condition. 

\textbf{Computation services} include off-loading either a computation task to a road-side unit or other cars in the neighborhood. We propose two computation models in such a scenario. The first model consists of a set of predefined computation tasks for a set of well-known V2X applications. For example, a vehicle requiring the object recognition computation service can send the image frame and the task name associated with the object recognition to the vehicle providing the computation service, which in return would send the objects present in the frame. Creating a set of standard computing tasks for the known application scenarios would not only reduce the communication overhead but also increases the efficiency of the off-loading process. The second model enables any MOTIVE instance to off-load any computation task to neighboring vehicles or road-side units. This model follows our prior work, SmartEdge~\cite{wright2018smartedge}, which supports the execution of arbitrary computation tasks. The challenges and design trade-offs of computation-offloading are discussed in Section~\ref{sec:rc}.


\textbf{Trust and Ratings:} The services provided and consumed through MOTIVE instances help drivers to make an informed decision or to gain monetary benefits. Vehicles and road-side units running the MOTIVE instance may behave dishonestly, wherein the devices may deliberately share incorrect information about the contact duration by providing malicious information or violate the service agreement by abruptly leaving the neighborhood after receiving the payment for the service. To identify bad actors in the MOTIVE eco-system, we propose a permissioned setup, wherein each user is required to register their driving license and plate information to the MOTIVE platform and infrastructure nodes also must register in an identified manner. After each transaction, each user is rated based on their behavior by the other party. Users are required to maintain a certain rating to participate in the MOTIVE transaction. When the rating of the device drops below a certain threshold, the user may be removed from the platform or subject to other penalties.

\textbf{Link Prediction:} The link prediction algorithm estimates the contact duration for the vehicle before providing or consuming services. This algorithm uses the navigation data, traffic situation, distance between the vehicles, wireless communication range of the radio, and the speed limit for estimating the duration for which the two vehicles will stay in contact. Our algorithm maintains the minimal contact duration needed to provide or consume each service in its registry. Thus, the scheduling of services is done when the contact duration is sufficient for the requested services. \textbf{Resource Management and Scheduling:} Each MOTIVE instance is capable of providing computation and storage resources for other MOTIVE instances in the neighborhood. The scheduling of resources for a given service depends on the contact duration, rating of the user, and the account status. When the scheduling criteria are met, the MOTIVE instance handles the service request by allocating the necessary computation, storage, and communication resources.



\section{Research Challenges and Design Choices}
\label{sec:rc}
\subsection{Resource Description and Discovery: how to describe the resources and services}
The topic-based publish-subscribe communication model uses topics to label the name of the data source. A standard set of topics can be developed to build V2X applications~\cite{5763516}. For example, the topic "speed" can be created to announce the current speed of the vehicle. For compute services, one approach is to have named functions (similar to AWS lambda~\cite{poccia2016aws}) that are available at the compute end e.g. “objectDetection(inputImage, objectName)” could take an image as an input and return yes/no if the named object is in the image, more like a function as a service (FaaS) model. At the other extreme, the compute end specifies a processor type and availability and the consumer may send arbitrary job and the input data, more like infrastructure as a service (IaaS) model. 


\subsection{Link Prediction: why it is needed}
Quynh \emph{et al.}~\cite{10.1007/978-3-319-75683-7_13} show that the contact duration of vehicles change based on the location, time, and the direction of the vehicle using real trace data from taxis in Shanghai. It is essential to identify the necessary parameters for calculating the contact duration to increase transaction reliability. 


\subsection{Prioritization and Scheduling}
MOTIVE enables the vehicle and infrastructure owners to schedule services to maximize their incentive. It is important to identify what metrics should be used for scheduling. A number of short jobs may provide the same incentive as the one long compute-intensive tasks. The question is how to schedule service based on the contract duration, incentive, and the application-specific metrics. Another question is regarding the relationship between the priority and the reputation, as each MOTIVE instance can choose a policy that maximizes their reward, but it may come at the cost of rejecting the requests of one or more vehicles.


\subsection{Trusted Data Delivery and Payment}
The problem with peer-to-peer data and payment is the classic ``Buyer and Seller's dilemma~\cite{DBLP:journals/corr/abs-1806-08379}", wherein the seller may leave after accepting the payment without providing the service, or the buyer may receive the service and leave without paying for the service. Asgaonkar \emph{et al.}~\cite{DBLP:journals/corr/abs-1806-08379} propose double escrow smart contract as a solution to ensure trusted data and payment delivery between the participants. Radhakrishnan \emph{et al.}~\cite{radhakrishnanstreaming} propose SDPP, which is a streaming data payment protocol based on a moving window that enables the parties to exchange data and payment in multiple steps. Such approaches can be incorporated into MOTIVE for reliable service and payment transactions.


\subsection{Verification of compute when offloading}
When off-loading computation to remote devices, it is critical to ensure that the remote device performed the computation. The device that off-loaded cannot redo the entire computation to ensure correctness. An approach is needed to reliably and efficiently off-load and verify computation. Wright \emph{et al.}~\cite{wright2018smartedge} advocate embedding a set of inputs with known outputs as part of the computation task, and verify only the outputs for the known inputs. This ensures that the computation was performed correctly by the remote device as long as the remote device is not aware of the positions of the embedded inputs.


\subsection{Identity, Ratings and Reputation}
We advocate using driver ID or other means of securing identity for each MOTIVE device in a permissioned set up, whether or not implemented on a public or private blockchain. Ratings should be vetted, and only different parties should be allowed to rate each other. At what granularity to implement ratings and reputation (sub-transactions vs. interactions) is an open question. 

\subsection{Quality of service}
Cloud-backup could be provided optionally as a way to provide reliability, so that data or compute results can be delivered even when vehicles are out of range but at the expense of higher latency and cost. Such a solution can be considered, but it has several trade-offs since cloud compute, storage and cellular bandwidth are all expensive resources. 

\subsection{Data and Compute Pricing}
The pricing for the data and computation should be market determined, or in some cases, it may make sense to do a dynamic auction, but that could be too complex for short interactions. Identifying models for data and compute pricing is vital to creating a market. Harsha \emph{et al.}~\cite{6880417} present a dynamic model for energy management problem, which could be adapted to price data and computation. 




\section{Preliminary Implementation and Early Results}
\label{sec:pi}
\begin{figure}[t]
\begin{center}
    \centerline{\includegraphics[height=3in,width=3in, keepaspectratio]{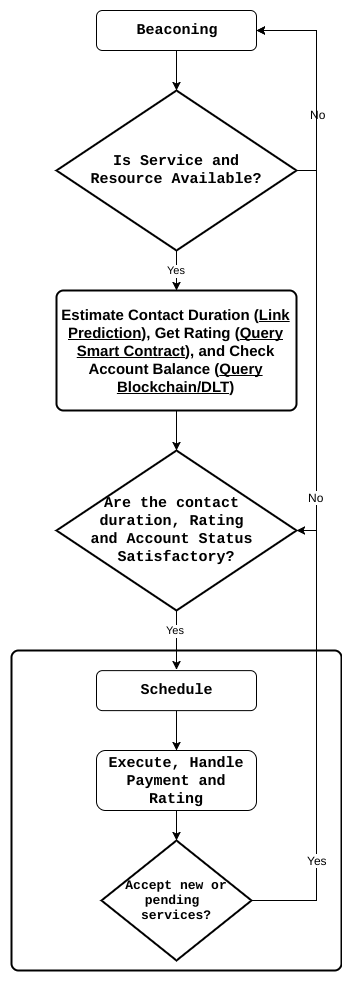}}
    \caption{MOTIVE execution flow.}
    \label{fig:flow}
\end{center}
\end{figure}

Figure~\ref{fig:flow} shows the execution flow of MOTIVE. Our preliminary implementation is carried out on a laptop machine and a Raspberry Pi device. We used WiFi as the V2X communication channel for device-to-device communication and beaconing. A secondary USB WiFi Dongle was used for Internet connection.

We used the WiFi's AdHoc feature for \textbf{beaconing} and peer-to-peer networking. Whenever a MOTIVE device comes in contact with another MOTIVE device, the devices establish a link and exchange the provided and required services. The \textbf{ratings} smart contract in Ethereum, which consists of three functions to add a new user to the MOTIVE eco-system, get the rating of an existing user, and rate an user after a transaction. We implemented a client application for handling the \textbf{payment and recording} the transactions using IOTA. Our proof-of-concept implementation and the demonstration videos are available at \url{https://github.com/ANRGUSC/MOTIVE}. Link prediction, scheduling, and dashboard functionality are being implemented.

The MOTIVE framework is agnostic to the underlying blockchain platform although our proof-of-concept implementations are carried out on particular platforms, namely IOTA and Ethereum. In its present form, MOTIVE technically only requires blockchain platform with support for rating, record, and payment functionalities (and not necessarily a full Turing-complete smart contract environment as provided by Ethereum).
\section{Conclusion}
\label{ref:con}
In this work, we have presented MOTIVE, a platform with support for providing and consuming data and computing services in return for micropayments. Besides, the link prediction algorithm of MOTIVE allows the vehicle to estimate the contact duration before scheduling the services. The rating mechanism used in MOTIVE allows the vehicle to verify the reputation of other vehicles for a trustworthy peer-to-peer transaction. Lastly, the article has presented the research challenges and design choices when developing a platform for V2X applications based on blockchain and distributed ledger technologies.

\bibliographystyle{IEEEtran}
\bibliography{references}

\begin{thebibliography}{10}
\providecommand{\url}[1]{#1}
\csname url@samestyle\endcsname
\providecommand{\newblock}{\relax}
\providecommand{\bibinfo}[2]{#2}
\providecommand{\BIBentrySTDinterwordspacing}{\spaceskip=0pt\relax}
\providecommand{\BIBentryALTinterwordstretchfactor}{4}
\providecommand{\BIBentryALTinterwordspacing}{\spaceskip=\fontdimen2\font plus
\BIBentryALTinterwordstretchfactor\fontdimen3\font minus
  \fontdimen4\font\relax}
\providecommand{\BIBforeignlanguage}[2]{{%
\expandafter\ifx\csname l@#1\endcsname\relax
\typeout{** WARNING: IEEEtran.bst: No hyphenation pattern has been}%
\typeout{** loaded for the language `#1'. Using the pattern for}%
\typeout{** the default language instead.}%
\else
\language=\csname l@#1\endcsname
\fi
#2}}
\providecommand{\BIBdecl}{\relax}
\BIBdecl

\bibitem{wachenfeld2016use}
W.~Wachenfeld, H.~Winner, J.~C. Gerdes, B.~Lenz, M.~Maurer, S.~Beiker,
  E.~Fraedrich, and T.~Winkle, ``Use cases for autonomous driving,'' in
  \emph{Autonomous driving}.\hskip 1em plus 0.5em minus 0.4em\relax Springer,
  2016, pp. 9--37.

\bibitem{gerla2014internet}
M.~Gerla, E.-K. Lee, G.~Pau, and U.~Lee, ``Internet of vehicles: From
  intelligent grid to autonomous cars and vehicular clouds,'' in \emph{Internet
  of Things (WF-IoT), 2014 IEEE World Forum on}.\hskip 1em plus 0.5em minus
  0.4em\relax IEEE, 2014, pp. 241--246.

\bibitem{mwasilu2014electric}
F.~Mwasilu, J.~J. Justo, E.-K. Kim, T.~D. Do, and J.-W. Jung, ``Electric
  vehicles and smart grid interaction: A review on vehicle to grid and
  renewable energy sources integration,'' \emph{Renewable and sustainable
  energy reviews}, vol.~34, pp. 501--516, 2014.

\bibitem{siegel2018survey}
J.~E. Siegel, D.~C. Erb, and S.~E. Sarma, ``A survey of the connected vehicle
  landscape—architectures, enabling technologies, applications, and
  development areas,'' \emph{IEEE Transactions on Intelligent Transportation
  Systems}, vol.~19, no.~8, pp. 2391--2406, 2018.

\bibitem{Pedrosa:2018:CBT:3211933.3211949}
\BIBentryALTinterwordspacing
A.~R. Pedrosa and G.~Pau, ``Chargeltup: On blockchain-based technologies for
  autonomous vehicles,'' in \emph{Proceedings of the 1st Workshop on
  Cryptocurrencies and Blockchains for Distributed Systems}, ser.
  CryBlock'18.\hskip 1em plus 0.5em minus 0.4em\relax New York, NY, USA: ACM,
  2018, pp. 87--92. [Online]. Available:
  \url{http://doi.acm.org/10.1145/3211933.3211949}
\BIBentrySTDinterwordspacing

\bibitem{Leiding:2016:SBV:2968219.2971409}
\BIBentryALTinterwordspacing
B.~Leiding, P.~Memarmoshrefi, and D.~Hogrefe, ``Self-managed and
  blockchain-based vehicular ad-hoc networks,'' in \emph{Proceedings of the
  2016 ACM International Joint Conference on Pervasive and Ubiquitous
  Computing: Adjunct}, ser. UbiComp '16.\hskip 1em plus 0.5em minus 0.4em\relax
  New York, NY, USA: ACM, 2016, pp. 137--140. [Online]. Available:
  \url{http://doi.acm.org/10.1145/2968219.2971409}
\BIBentrySTDinterwordspacing

\bibitem{leiding2018enabling}
B.~Leiding and W.~V. Vorobev, ``Enabling the vehicle economy using a
  blockchain-based value transaction layer protocol for vehicular ad-hoc
  networks,'' 2018.

\bibitem{8455893}
Z.~Lu, Q.~Wang, G.~Qu, and Z.~Liu, ``Bars: A blockchain-based anonymous
  reputation system for trust management in vanets,'' in \emph{2018 17th IEEE
  International Conference On Trust, Security And Privacy In Computing And
  Communications/ 12th IEEE International Conference On Big Data Science And
  Engineering (TrustCom/BigDataSE)}, Aug 2018, pp. 98--103.

\bibitem{5gaa}
D.~Sabella and et. al, ``Toward fully connected vehicles: Edge computing for
  advanced automotive communications,'' 2017.

\bibitem{8442848}
J.~A.~L. Calvo and R.~Mathar, ``Secure blockchain-based communication scheme
  for connected vehicles,'' in \emph{2018 European Conference on Networks and
  Communications (EuCNC)}, June 2018, pp. 347--351.

\bibitem{bhargava2016optimizing}
A.~Bhargava, S.~Congero, T.~Ferrell, A.~Jones, L.~Linsky, J.~Mohan, and
  B.~Krishnamachari, ``Optimizing downloads over random duration links in
  mobile networks,'' in \emph{Computer Communication and Networks (ICCCN), 2016
  25th International Conference on}.\hskip 1em plus 0.5em minus 0.4em\relax
  IEEE, 2016, pp. 1--9.

\bibitem{6573913}
J.~Ahn, M.~Sathiamoorthy, B.~Krishnamachari, F.~Bai, and L.~Zhang, ``Optimizing
  content dissemination in vehicular networks with radio heterogeneity,''
  \emph{IEEE Transactions on Mobile Computing}, vol.~13, no.~6, pp. 1312--1325,
  June 2014.

\bibitem{wright2018smartedge}
K.-L. Wright, M.~Martinez, U.~Chadha, and B.~Krishnamachari, ``Smartedge: A
  smart contract for edge computing.''

\bibitem{5763516}
T.~Mishra, D.~Garg, and M.~M. Gore, ``A publish/subscribe communication
  infrastructure for vanet applications,'' in \emph{2011 IEEE Workshops of
  International Conference on Advanced Information Networking and
  Applications}, March 2011, pp. 442--446.

\bibitem{poccia2016aws}
D.~Poccia, \emph{AWS Lambda in Action: Event-driven serverless
  applications}.\hskip 1em plus 0.5em minus 0.4em\relax Manning Publications
  Co., 2016.

\bibitem{10.1007/978-3-319-75683-7_13}
Q.~Nguyen and B.~Krishnamachari, ``A study of contact durations for vehicle to
  vehicle communications,'' in \emph{Proceedings of International Symposium on
  Sensor Networks, Systems and Security}, N.~S. Rao, R.~R. Brooks, and C.~Q.
  Wu, Eds.\hskip 1em plus 0.5em minus 0.4em\relax Cham: Springer International
  Publishing, 2018, pp. 173--183.

\bibitem{DBLP:journals/corr/abs-1806-08379}
\BIBentryALTinterwordspacing
A.~Asgaonkar and B.~Krishnamachari, ``Solving the buyer and seller's dilemma:
  {A} dual-deposit escrow smart contract for provably cheat-proof delivery and
  payment for a digital good without a trusted mediator,'' \emph{CoRR}, vol.
  abs/1806.08379, 2018. [Online]. Available:
  \url{http://arxiv.org/abs/1806.08379}
\BIBentrySTDinterwordspacing

\bibitem{radhakrishnanstreaming}
R.~Radhakrishnan and B.~Krishnamachari, ``Streaming data payment protocol
  (sdpp) for the internet of things,'' 2018.

\bibitem{6880417}
P.~Harsha and M.~Dahleh, ``Optimal management and sizing of energy storage
  under dynamic pricing for the efficient integration of renewable energy,''
  \emph{IEEE Transactions on Power Systems}, vol.~30, no.~3, pp. 1164--1181,
  May 2015.

\end{thebibliography}

\end{document}